\documentclass[preprint]{aastex63}

\received{June 1, 2019}
\revised{August 18, 2019}
\accepted{\today}
\shorttitle{Damping of filament longitudinal oscillations}
\shortauthors{Zhang et al.}
\usepackage{amsmath,amssymb}
\usepackage{graphbox,graphicx}
\usepackage{enumitem}

\graphicspath{{figures/}}

\newcommand{\unit}[1]{\ensuremath{\,\mathrm{#1}}}
\def\mp{m_\mathrm{p}}
\def\kB{k_\mathrm{B}}
\def\nH{n_\mathrm{H}}
\def\ne{n_\mathrm{e}}

\def\tot{\mathrm{tot}}
\def\kms{$\rm{km~s}^{-1}$}

\def\ek{e_\mathrm{k}}

\begin{document}

\title{Damping Mechanisms of the Solar Filament Longitudinal Oscillations in Weak Magnetic Field}

\correspondingauthor{P. F. Chen}
\email{chenpf@nju.edu.cn}

\author{L. Y. Zhang}
\affil{School of Astronomy and Space Science, Nanjing University,
Nanjing 210023, China}
\affil{Key Laboratory of Modern Astronomy \& Astrophysics, Nanjing University, China}

\author{C. Fang}
\affil{School of Astronomy and Space Science, Nanjing University,
Nanjing 210023, China}
\affil{Key Laboratory of Modern Astronomy \& Astrophysics, Nanjing University, China}

\author[0000-0002-7289-642X]{P. F. Chen}
\affil{School of Astronomy and Space Science, Nanjing University,
Nanjing 210023, China}
\affil{Key Laboratory of Modern Astronomy \& Astrophysics, Nanjing University, China}

\begin{abstract}

Longitudinal oscillations of solar filament have been investigated via
numerical simulations continuously, but mainly in one dimension (1D), where 
the magnetic field line is treated as a rigid flux tube. Whereas those 
one-dimensional simulations can roughly reproduce the observed oscillation
periods, implying that gravity is the main restoring force for filament
longitudinal oscillations, the decay time in one-dimensional simulations
is generally longer than in observations. In this paper, we perform a
two-dimensional (2D) non-adiabatic magnetohydrodynamic simulation of
filament  longitudinal oscillations, and compare it with the 2D adiabatic case 
and 1D adiabatic and non-adiabatic cases. It is found that, whereas both 
non-adiabatic processes (radiation and heat conduction) can significantly 
reduce the decay time, wave leakage is another important mechanism to 
dissipate the kinetic energy of the oscillating filament when the magnetic 
field is weak so that gravity is comparable to Lorentz force. In this case,
our simulations indicate that the pendulum model might lead to an error of 
$\sim$100\% in determining the curvature radius of the dipped magnetic field 
using the longitudinal oscillation period when the gravity to Lorentz force 
ratio is close to unity.
\end{abstract}


\section{Introduction}

% - filaments
Solar filaments, or called prominences when appearing above the solar
limb, are cold dense plasma concentrations embedded in the hot
tenuous corona. In the early era when the spatial resolution of
observations was not high, filaments seem to be static, hence they were
thought to be held in equilibrium with the Lorentz force balancing the
gravity. Therefore, it was proposed that the local magnetic field of
filaments should have a dipped configuration, either of a normal-polarity type
\citep{kipp57} or inverse-polarity type \citep{kupe74}. After the dynamic 
features were discovered, e.g., the counterstreamings \citep{zirk98}, it was 
suggested that a dipped magnetic configuration is
not necessary, and a prominence could be be a completely dynamic
structure, with existing plasma draining down and new plasma replenishing it
\citep{karp01}. Such a possibility was validated in some observations
\citep{wang99, zou16}.

Even in the case with magnetic dips, some filaments are never static. They 
oscillate in response to any perturbations, which are ubiquitous in the solar 
atmosphere. From the physical point of view, filament oscillations can be 
divided into transverse and longitudinal modes \citep{shen14, zhan17, arre18}, 
where the oscillation direction is perpendicular and
parallel to the local magnetic field, respectively. Similar to other
oscillating phenomena, filament oscillations, characterized by oscillation 
period and decay time, can also be utilized to estimate some physical 
parameters, mainly the magnetic field. This is called prominence seismology 
\citep{arre18}. Hence, it is important to understand what
determines the oscillation period and what determines the decay time.

Filament longitudinal oscillations were first reported by \citet{Jing2003}, 
and in fact, the counterstreamings existing in many solar filaments might be 
due to longitudinal oscillations of filament threads \citep{lin03}, although 
the counterstreamings in some filaments are alternating unidirectional flows 
\citep{zou17} or are the combination of filament thread longitudinal 
oscillations and 
unidirectional flows \citep{chen14}. Several mechanisms were put forward to 
account for the restoring force of the filament longitudinal oscillations, 
such as the magnetic field-aligned component of gravity, gas pressure, and 
magnetic tension force \citep{Jing2003, vrsn07}. With one-dimensional (1D) 
hydrodynamic simulations, \citet{Luna2012b} and \citet{zhan12} verified 
that the field-aligned component of gravity can explain the observed periods, 
which are around 1 hour. In particular, the magnetic configuration in
\citet{zhan12} was extracted from high-resolution observations, the agreement 
of the oscillation period between the simulations and the observations is 
strongly indicative of that the field-aligned component of gravity is the 
restoring force for the filament longitudinal oscillations, i.e., the filament 
longitudinal oscillations can be accounted for with the pendulum model. Hence, 
the longitudinal oscillation period of solar filaments was considered to be 
able to diagnose the curvature of the dipped magnetic field, an important part 
of prominence seismology.

It is noted that the above-mentioned pendulum model is somewhat simplified,
where the magnetic field is assumed to be a rigid flux tube. In real
observations, the magnetic field might be deformed by the moving filaments 
\citep{li12}. The deformed magnetic field results in change of the 
field-aligned gravity, which then alters the oscillation period. The 
deformation would make the prominence seismology not so straightforward.
\citet{Luna2016} performed two-dimensional (2D) magnetohydrodynamic (MHD) 
simulations of filament longitudinal
oscillations, where they found that the magnetic field is
slightly deformed by the oscillating filament. It was suggested that the
deformation was so small that the oscillation period is similar to that
estimated from the simplified pendulum model. However, their result
might be due to that their magnetic field is not weak enough. In order
to quantify whether the filament gravity can significantly deform the
shape of the magnetic field, \citet{Zhou2018} defined a dimensionless
parameter, plasma $\delta={{\rho gL}\over {B^2/2\mu_0}}=11.5{n \over 
{10^{11}~\rm{cm}^{-3}}} {L \over {100~\rm{Mm}}} ({B \over {10~\rm{G}}})^{-2}$,
where $n$ is the number density of the prominence,
$L$ is the length of the prominence thread, and $B$ is the magnetic
field. If $\delta$ is much smaller than unity, the deformation of the
magnetic field would be small; If $\delta$ is comparable to or larger
than unity, the deformation of the magnetic field would be significant.
Based on the parameters used in \citet{Luna2016}, we found that their
$\delta$ is $\sim$0.2. Hence, it is expected that the oscillating
filament would not modify the magnetic field significantly, hence the
pendulum model should be valid. In this paper, we plan to extend the
$\delta$ parameter to be around unity, a regime where magnetic field
would be deformed against the filament gravity, and to investigate
whether the pendulum model still works fine.

More importantly, although
\citet{zhan12} successfully reproduced the observed oscillation period
of a prominence, the decay time of the oscillation in their simulation
is 1.5 times larger than in the observations. They attributed the longer
decay time in their simulations to the absence of other energy loss
mechanisms, such as wave leakage, which can be modeled only in 2D or 3D
MHD simulations. Therefore, in this paper, we also intend to
investigate whether the decay time of filament oscillations in 2D would
be significantly reduced in contrast to the corresponding 1D case.

This paper is organized as follows: The numerical method is described in
\S\ref{sec:method}, the numerical results are presented in
\S\ref{sec:res}, which are followed by discussions in \S\ref{sec:dis}. A
summary is given in \S\ref{sec:sum}.

%%%%%%%%%%%%%%%%%%%%%%%%%%%%%%%%%%%%%%%%%%%%%%%%%%%%%%%%%%%%%%%%%%%%%%%%%%%%%%
\section{Numerical Method} \label{sec:method}

We solve the following 2D ideal MHD equations in the $x$-$z$ plane to investigate the dynamics of filament oscillations in this work:
%-----------------------------------------------------------------------
%\begin{gather}
\begin{equation}
    \frac{\partial \rho}{\partial t} + \nabla \cdot (\rho \boldsymbol{v}) = 0\,, \label{eq:rho}
\end{equation}

\begin{equation}
    \frac{\partial(\rho \boldsymbol{v})}{\partial t}
        + \nabla \cdot \left(\rho \boldsymbol{v}\boldsymbol{v} - \frac{\boldsymbol{B}\boldsymbol{B}}{\mu_0}\right)
        + \nabla \left(p+\frac{\boldsymbol{B}^2}{2\mu_0}\right)
        = -\rho \boldsymbol{g}\,, \label{eq:mom}
\end{equation}

\begin{equation}
    \frac{\partial e_\tot}{\partial t} + \nabla\cdot\left(e_\tot \boldsymbol{v}
        +\left(p+\frac{\boldsymbol{B}^2}{2\mu_0}\right)\boldsymbol{v}
        - \frac{\boldsymbol{B}\boldsymbol{B}}{\mu_0}\cdot\boldsymbol{v}\right)
        = -\rho \boldsymbol{v} \cdot \boldsymbol{g} + \nabla\cdot{\boldsymbol{q}} - {\ne}{\nH}\Lambda (T) + H(z)\,, \label{eq:eng}
\end{equation}

\begin{equation}
    \frac{\partial\boldsymbol{B}}{\partial t} + \nabla\cdot(\boldsymbol{v}\boldsymbol{B}-\boldsymbol{B}\boldsymbol{v}) = 0\,, \label{eq:b}
%\end{gather}
\end{equation}
%-----------------------------------------------------------------------
where $\rho = 1.4 \mp \nH$ is the mass density,
$\boldsymbol{v}=(\boldsymbol{v}_x$, $\boldsymbol{v}_z)$ is the
velocity, $\boldsymbol{B}=(\boldsymbol{B}_x$, $\boldsymbol{B}_z)$ is
the magnetic field, $p = 2.3 \nH \kB T$ is the thermal pressure ($\mp$
is the proton mass and $\kB$ is the Boltzmann constant), and
$e_\tot=\rho v^2/2 + p/(\gamma - 1)+\boldsymbol{B}^2/(2\mu_0)$ is the total
energy density, where $\gamma = 5/3$ is the ratio of the specific heats.
% - gravity
The gravity is set to be uniform, i.e.,
$\boldsymbol{g} = g_\odot \hat{\boldsymbol{e}}_z$, where $g_\odot$ is $274\unit{m\,s^{-2}}$.
% - heating sources
In the energy equation, i.e., Equation (\ref{eq:eng}), we consider thermal
conduction, the optically thin radiative cooling, and the background
heating. The field-aligned heat conduction is described as follows,
%-----------------------------------------------------------------------
\begin{equation}
    \boldsymbol{q} = \kappa_\parallel (\boldsymbol{b} \cdot \nabla{T})\,\boldsymbol{b},
    \label{eq:q}
\end{equation}
%-----------------------------------------------------------------------
where $\boldsymbol{q}$ is the heat flux vector, $\boldsymbol{b}=\boldsymbol{B}/|B|$ is the unit vector along the magnetic field,
$\kappa_{\parallel} = 10^{-6} T^{5/2}\unit{erg}\unit{cm}^{-1}\unit{s}^{-1}\unit{K}^{-1}$ is the Spitzer heat conductivity.
% - cooling
$\Lambda(T)$ is the radiative loss coefficient for the optically thin emission, which 
is obtained by interpolating a cooling table based on the SPEX package 
\citep[see][for details]{Schure2009}. The radiative loss coefficient is set 
to 0 when $T < 8\times 10^3\unit{K}$.
% - heating
$H(z) = H_0 e^{-z/H_m}$ is the steady heating term, where the amplitude
$H_0 = 1.5\times10^{-4}\unit{erg}\unit{cm}^{-3}\unit{s}^{-1}$ and
$H_m = 40\unit{Mm}$ is the scale height. This term is introduced in order to maintain the background corona.
%--------------------------------------- code
The MHD equations are numerically solved with the MPI Adaptive Mesh Refinement
Versatile Advection Code \citep[MPI-AMRVAC;][]{Keppens2012,Porth2014,Xia2018}.
In particular, the heat conduction term in Equation (\ref{eq:eng}) is solved with
an implicit scheme in order to avoid too small time steps.
% - region
The computational domain is a rectangular region in the range of the Cartesian coordinates
$x\in[-100, 100]\unit{Mm}$ and $z\in[0, 100]\unit{Mm}$.
The resolution of the finest mesh layer is 156 \unit{km}.

The simulated normal polarity filament is placed in a dipped magnetic field.
Following \citet{Luna2016}, we adopt a quadrupolar magnetic
configuration, which is described by $B_x=B_0\;(\cos
k_1x\;e^{-k_1(z-z_0)}-\cos k_2x\;e^{-k_2(z-z_0)})$ and $B_z=B_0\;(-\sin
k_1x\;e^{-k_1(z-z_0)}+\sin k_2x\;e^{-k_2(z-z_0)})$. Here, we take
$k_1=k_2/3=\pi/200$ Mm$^{-1}$, $z_0=4\unit{Mm}$, and $B_0=10\unit{G}$.
This configuration has a bald patch structure near the magnetic neutral line $x=0$.
The bottom boundary is
treated to be a line-tied one, with all quantities being fixed, the top
boundary is a free one, and the reflecting conditions are set on the two
side boundaries.
Similar to \citet{Zhou2018}, the initial conditions are realized via the following two steps:
\begin{enumerate}[label=(\arabic*)]
\item To set up a quiet Sun atmosphere: The temperature distribution
from the photosphere to the bottom of the corona is prescribed as
follows:
%-----------------------------------------------------------------------
\begin{equation}
    T(z)=
        \begin{cases}
        {T_{ph}} + ({T_{co}} - {T_{ph}})(1 + \tanh (z - {h_{tr}} - {c_1})/{w_{tr}})/2 & z \le h_{tr}, \\
        {(7{F_c}(z - {h_{tr}})/(2\kappa ) + {T_{tr}}^{7/2})^{2/7}} & z > h_{tr}, \\
    \end{cases}
    \label{eq:T}
\end{equation}
%-----------------------------------------------------------------------
\noindent
where the height of our transition region, $h_{tr}$, is set to a value of 
$1.5\unit{Mm}$, and its thickness, $w_{tr}$, is taken to be $0.2\unit{Mm}$. 
The temperatures of the photosphere, transition region, and the corona are 
$T_{ph}=9\times 10^3\unit{K}$, $T_{tr}=1.6\times 10^5\unit{K}$, and $T_{co}=1.5\times 
10^6\unit{K}$. With a bottom number density of $1.2\times 10^{14}\unit{cm^{-3}}$, the 
density distribution is calculated based on the hydrostatic equilibrium, where 
gravity is balanced by the gas pressure gradient. Such analytical distributions, 
together with the quadrupolar magnetic field described above, evolve 
gradually, and reach a real equilibrium state after $\sim$110 minutes in the MHD 
simulations, when the vertical temperature distribution is shown in Figure 
\ref{fig:Te-z}.

%-----------------------------------------------------------------------
\begin{figure}[!ht]
    \centering
    \includegraphics[width=.8\textwidth]{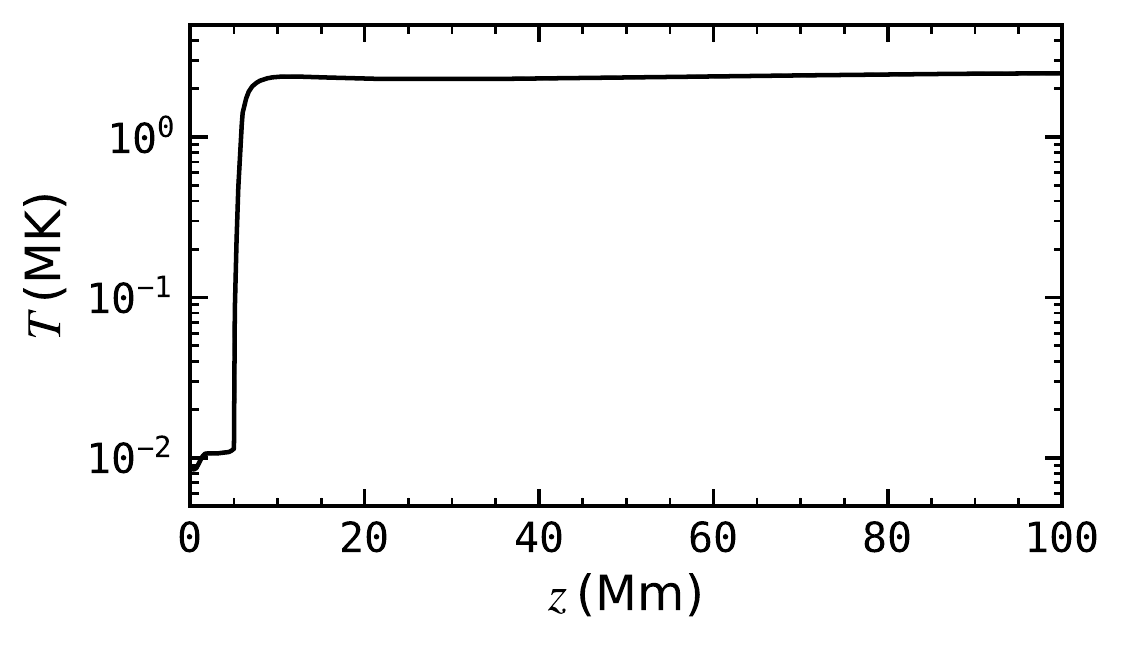}
    \caption{The vertical distribution of the temperature before the filament 
	is introduced.}
    \label{fig:Te-z}
\end{figure}
%-----------------------------------------------------------------------

\noindent
\item We place a bulk of dense plasma around the magnetic dips to represent a 
filament thread, whose density profile takes the following form:
%-----------------------------------------------------------------------
\begin{equation}
    \rho = \rho_{co} + \delta{\rho}\,e^{-\frac{(x-x_0)^4}{w_x^4}-\frac{(z-z_0)^4}{w_z^4}},
    \label{eq:rho1}
\end{equation}
%-----------------------------------------------------------------------
where $(x_0, z_0) = (0, 19.8)\unit{Mm}$ is the position of the mass center, 
$w_x = 4\unit{Mm}$, $w_z = 3\unit{Mm}$; $\rho_{co}$ is the plasma density in
the ambient corona obtained from Step (1) and $\delta{\rho} =99\,\rho_{co}$. 
The corresponding temperature, $T$, is also modified so that $\rho T$ remains 
unchanged. We then perform MHD simulations again.
\end{enumerate}
%-----------------------------------------------------------------------
%-----------------------------------------------------------------------
\begin{figure}[ht!]
    \centering
    \plotone{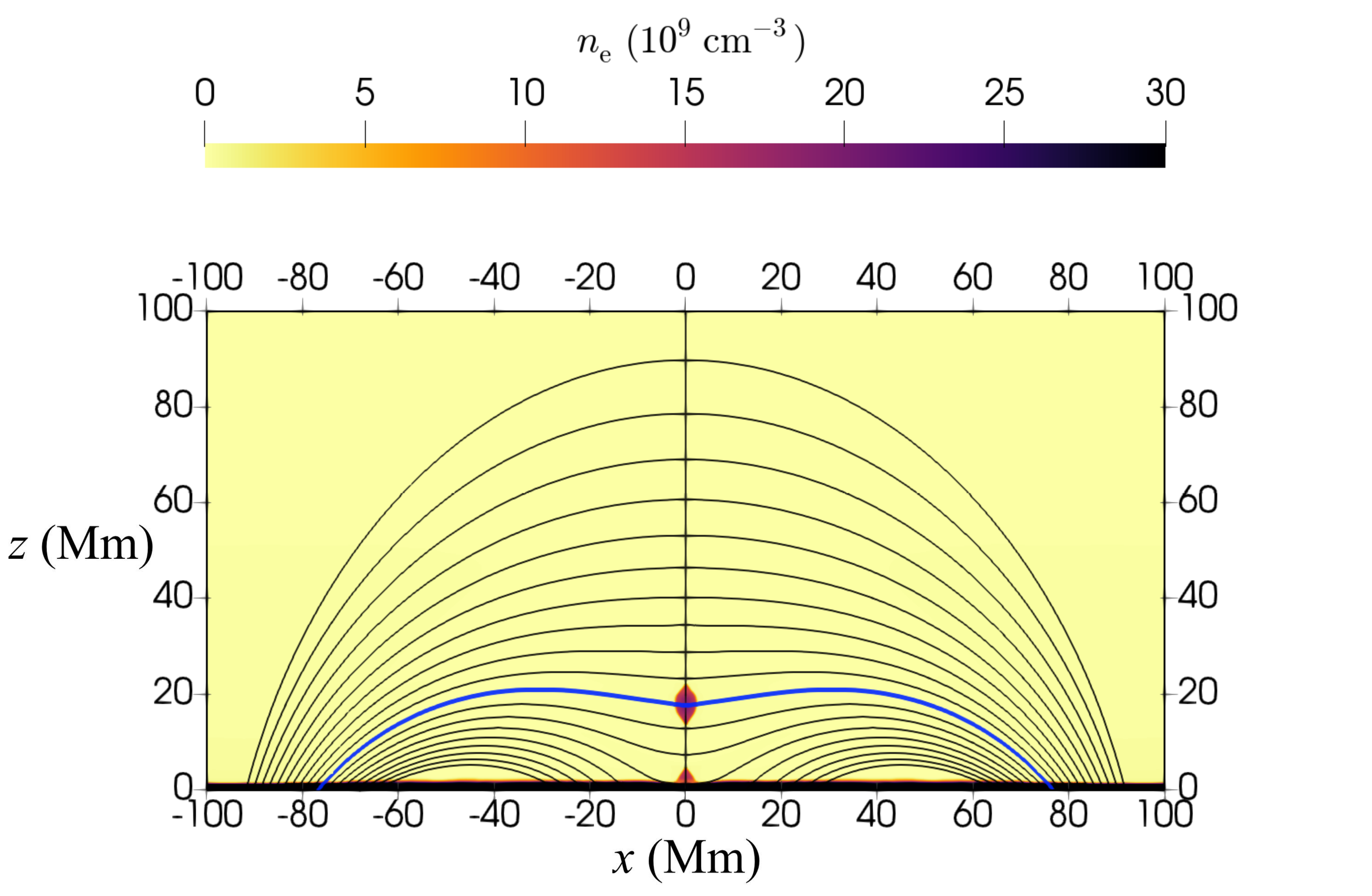}
	\caption{The distributions of the magnetic field (black lines) and 
	the plasma density (color scale) used as the initial conditions for 
	this paper. The blue line marks the single magnetic field line 
	threading the centroid of the filament.}
    \label{fig:init}
\end{figure}
%-----------------------------------------------------------------------
It is seen that the whole system evolves gradually. While plasmas is sucked 
into the filament thread due to thermal instability, the filament oscillates 
with an initial amplitude of 16 \kms. After $\sim$290 minutes, the amplitude of 
the filament transverse oscillation is still around 0.2 km s$^{-1}$. In order 
to speed up the decay, we set the velocity to be 0 everywhere in the simulation 
domain when the filament is at its equilibrium position. As a test, we continue 
the simulation and find the residual maximum velocity is only 0.06 km s$^{-1}$. 
Now, the system approaches a new equilibrium state, where the filament becomes 
$\sim$5.5$\unit{Mm}$ long in the $x$-direction and $\sim$10\unit{Mm} thick in 
the $z$-direction. The center of mass drops to $z_c=17.7\unit{Mm}$. The 
distributions of the magnetic field and plasma density are displayed in Figure 
\ref{fig:init} as solid lines and color scales, respectively. Such a state is 
taken to be the real initial conditions for the simulation work to be presented 
in this paper. Note that there is a second plasma condensation near the 
original point as shown in Figure \ref{fig:init}, which is formed due to 
thermal instability, and is not relevant to the study in this paper.

% - perturbation
According to \citet{Zhang2013}, the oscillation properties do not depend on
the type of perturbations, no matter it is due to impulsive momentum or 
localized heating. In this work we choose the former one. Similar to 
\citet{Luna2016}, the filament is perturbed with a magnetic
field-aligned velocity, which is expressed as
%-----------------------------------------------------------------------
\begin{equation}
    \boldsymbol{v}_0 = v_0\boldsymbol{b}\,e^{-\frac{(x-x_0)^4}{\tilde{w}_x}-\frac{(z-z_0)^4}{\tilde{w}_z}},
    \label{eq:v1}
\end{equation}
%-----------------------------------------------------------------------
where $v_0=20$ \kms and $\boldsymbol{b}$ is the same as in Equation 
(\ref{eq:q}), while $\tilde{w}_x=15\unit{Mm}$ and $\tilde{w}_z=10\unit{Mm}$. 
The ensuing evolution is calculated with the MPI-ARMVAC code. It is noted here 
that we also simulate adiabatic cases for comparison, where the radiation and 
heat conduction in Equation (\ref{eq:eng}) are removed.

%%%%%%%%%%%%%%%%%%%%%%%%%%%%%%%%%%%%%%%%%%%%%%%%%%%%%%%%%%%%%%%%%%%%%%%%%%%%%%
\section{Results} \label{sec:res}

\subsection{2D non-adiabatic case} \label{sec31}

%-----------------------------------------------------------------------
\begin{figure*}[!htp]
    \centering
    \includegraphics[width=.8\textwidth]{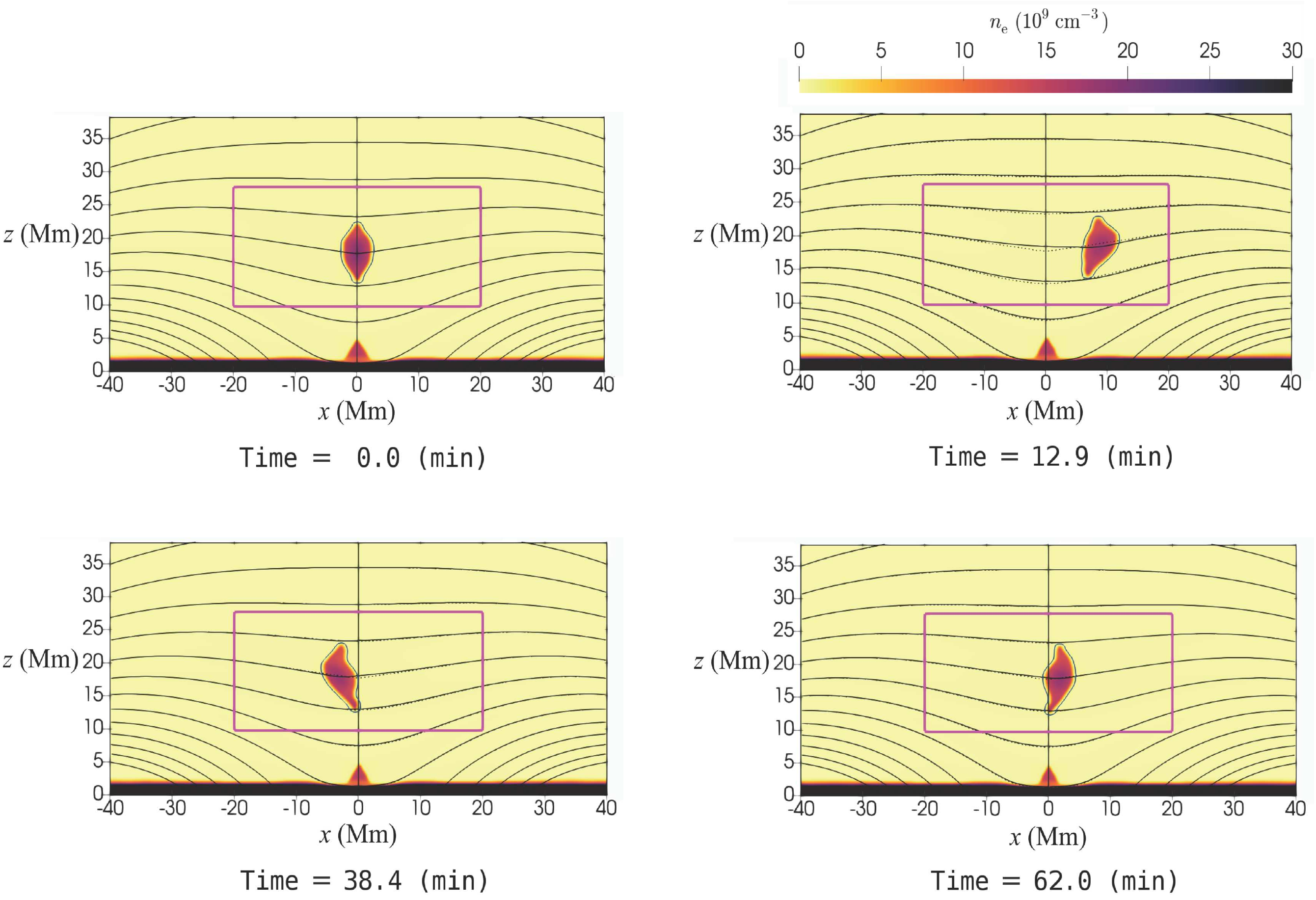}
	\caption{Evolution of the magnetic field (solid lines) and the plasma 
	density (color scale). For comparison, the initial magnetic field is 
	overplotted as the dashed lines.}
    \label{fig:evo}
\end{figure*}
%-----------------------------------------------------------------------

In this subsection, we present the numerical results in the 2D non-adiabatic 
case. The evolution of the filament oscillation is depicted in Figure 
\ref{fig:evo}, where the color represents the density and the solid lines 
correspond to the evolving magnetic field. To illustrate how the magnetic 
field is deformed, the initial magnetic field is overplotted as the dotted 
lines. It is seen that, after the velocity perturbation is imposed on the 
filament, the filament begins to oscillate, but the amplitude decays rapidly. 
It is also noticed that the upper and lower parts of the filament do not 
oscillate in phase since they have different oscillation periods, which are 
determined by different curvature radii of the local magnetic field \citep[see 
also][]{Luna2016}.  As a result, the filament changes its shape continuously.

%-----------------------------------------------------------------------
\begin{figure}[ht!]
    \centering
	\includegraphics[width=8cm,height=4cm]{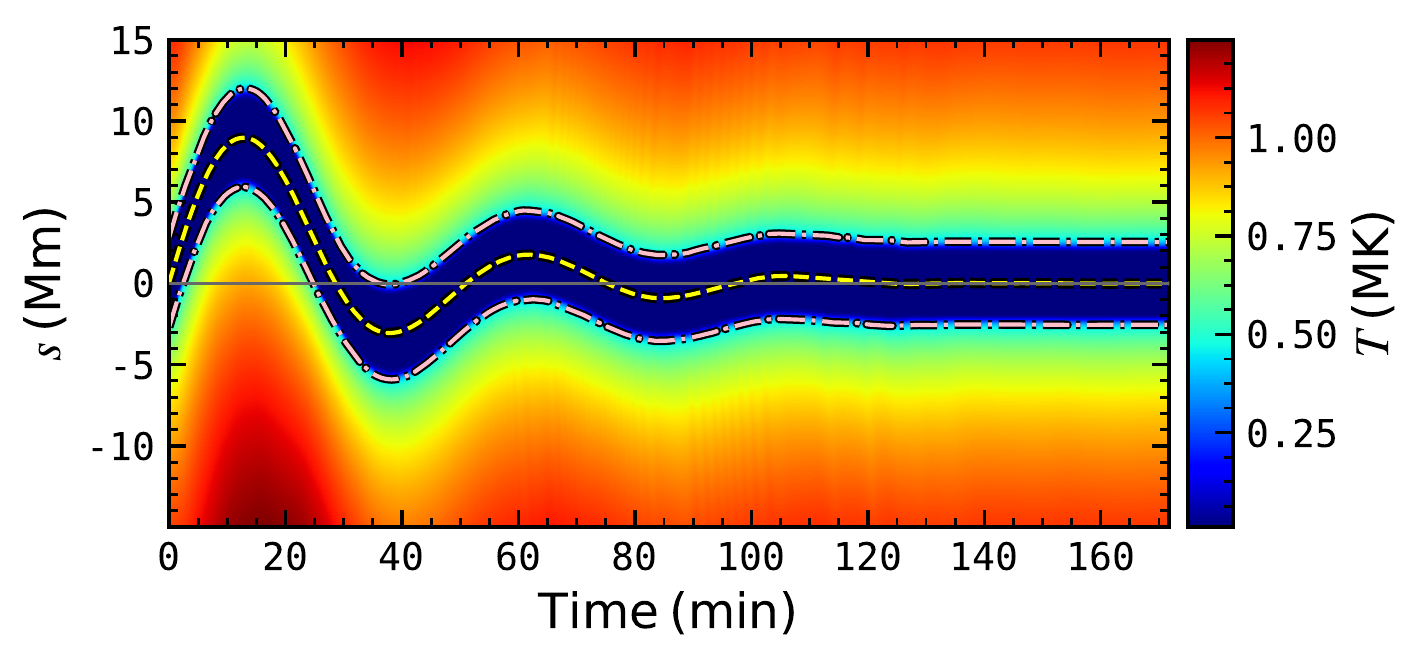}\hspace{1cm}
	\includegraphics[width=5cm,height=3.8cm]{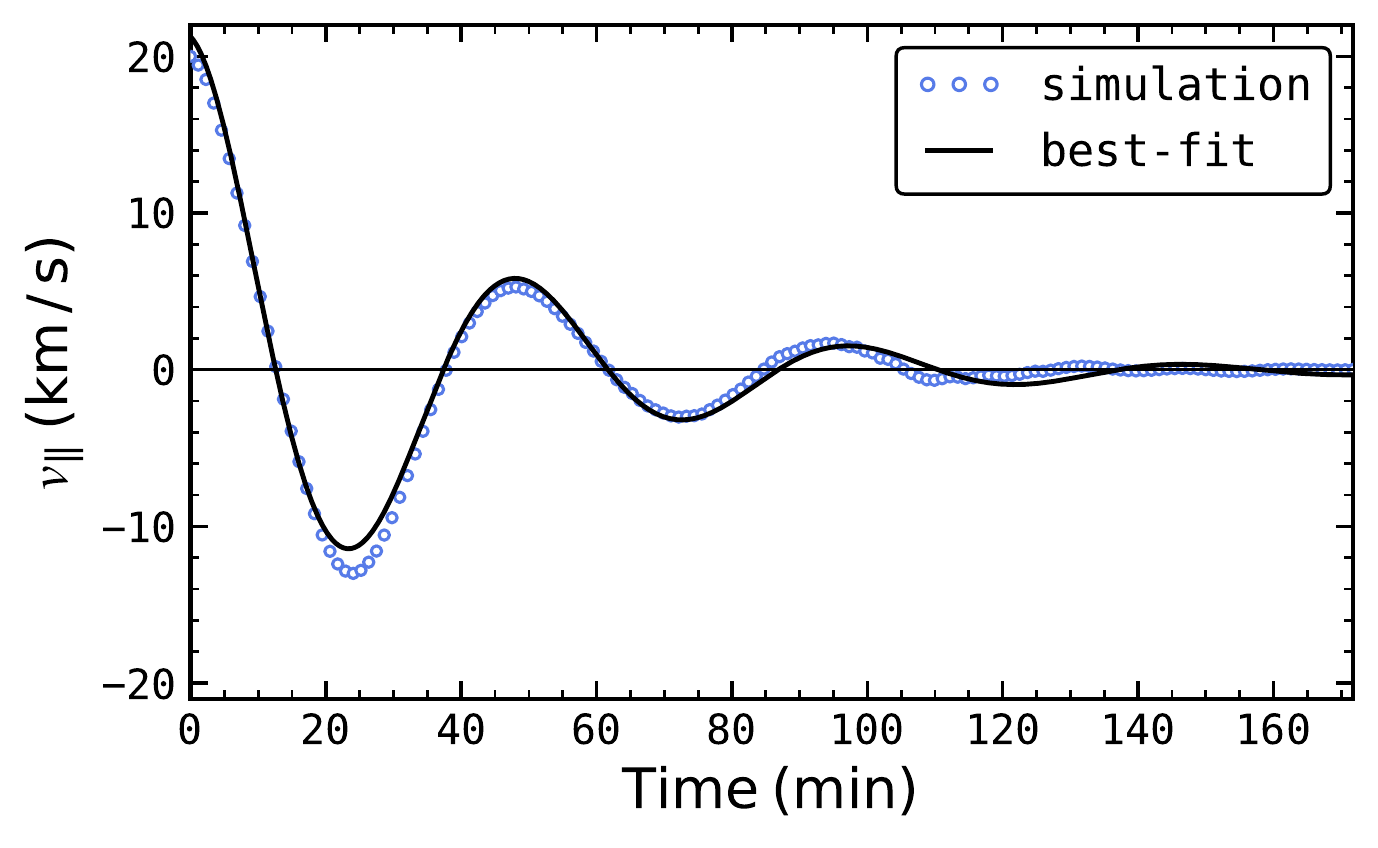}
    \caption{Left: Time-distance diagram of of the temperature distribution 
	along the magnetic field line threading the filament centroid. Right: 
	Evolution of the field-aligned velocity of the filament centroid.}
    \label{fig:Te-s}     %%% Fig. 4
\end{figure}
%-----------------------------------------------------------------------

In order to analyze the oscillation behavior more quantitatively, we extract 
the magnetic field line that passes through the initial centroid of the 
filament, which is indicated by the blue line in Figure \ref{fig:init}. This 
field line is rooted at the position with $x=\pm 76.8$ Mm and $z=0$. The 
time-distance diagram of the temperature distribution along this line is 
plotted in the left panel of Figure \ref{fig:Te-s}, where the origin of the 
distance is set at $x=0$. The filament corresponds to the blue area, whose 
center is indicated by the yellow dashed line, and whose boundaries are marked 
by the two dot-dashed lines. The filament velocity along the field line 
($v_{\parallel}$) is 
determined by the velocity of the filament centroid, and its evolution is 
displayed in the right panel of Figure \ref{fig:Te-s} as the blue circles. It 
has the typical damped sinusoidal profile. Therefore, we fit $v_{\parallel}$ 
with the following damped sine function via the least-square method,

%-----------------------------------------------------------------------
\begin{equation}
    v_{\parallel} = v_0 e^{-t/\tau} \sin\big(2\pi t/P + \varphi\big),
    \label{eq:fit0}
\end{equation}
%-----------------------------------------------------------------------
where $v_0$ is the initial amplitude, $P$ is the oscillation period, $\tau$ is 
the decay time, and $\varphi$ is the phase angle. It turns out that $v_0=20$ 
\kms, $P=49$ minutes, $\tau=38$ minutes, and $\varphi=\pi/2$. The fitted curve 
is overplotted in the right panel of Figure \ref{fig:Te-s} as the black line. 
It is seen that the fitting is reasonable for the first 1.5 periods, and the 
deviation becomes remarkable after that. It is evident that the oscillation 
period should be shorter and shorter, rather being a constant, in the late 
stage of the evolution.

\subsection{1D non-adiabatic case}

In order to find out what new effects are brought by the two dimensions,
we perform a 1D non-adiabatic hydrodynamic simulation for comparison. In
order to make the comparison meaningful, the magnetic configuration in
the 1D case is taken from the initial magnetic field line across the
filament centroid, i.e., the blue line in Figure \ref{fig:init}. Similar
to the initial condition-producing procedure for the 2D simulation, a
segment of filament is inset into the magnetic dip, with the length and
density identical to the 2D case.
%-----------------------------------------------------------------------
\begin{figure}[ht!]
    \centering
	\includegraphics[align=t,width=8cm,height=4.5cm]{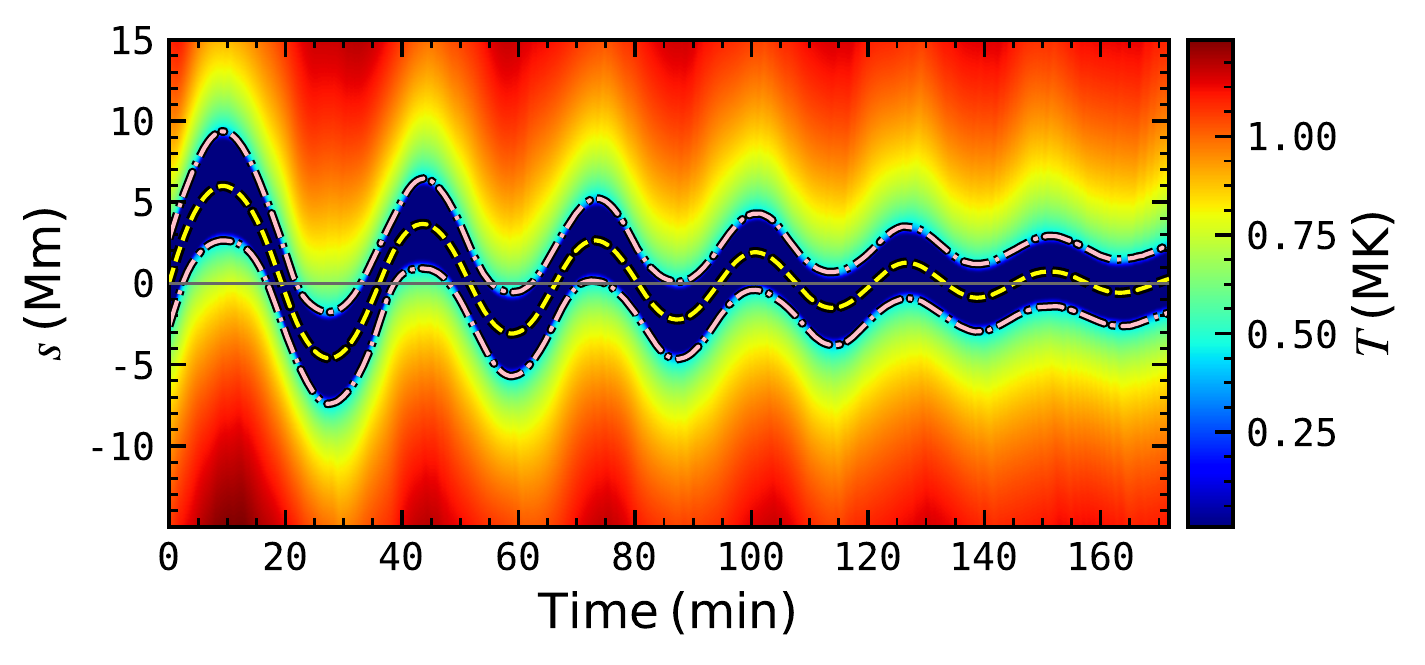}\hspace{1cm}
    \includegraphics[align=t,width=6cm,height=4.3cm]{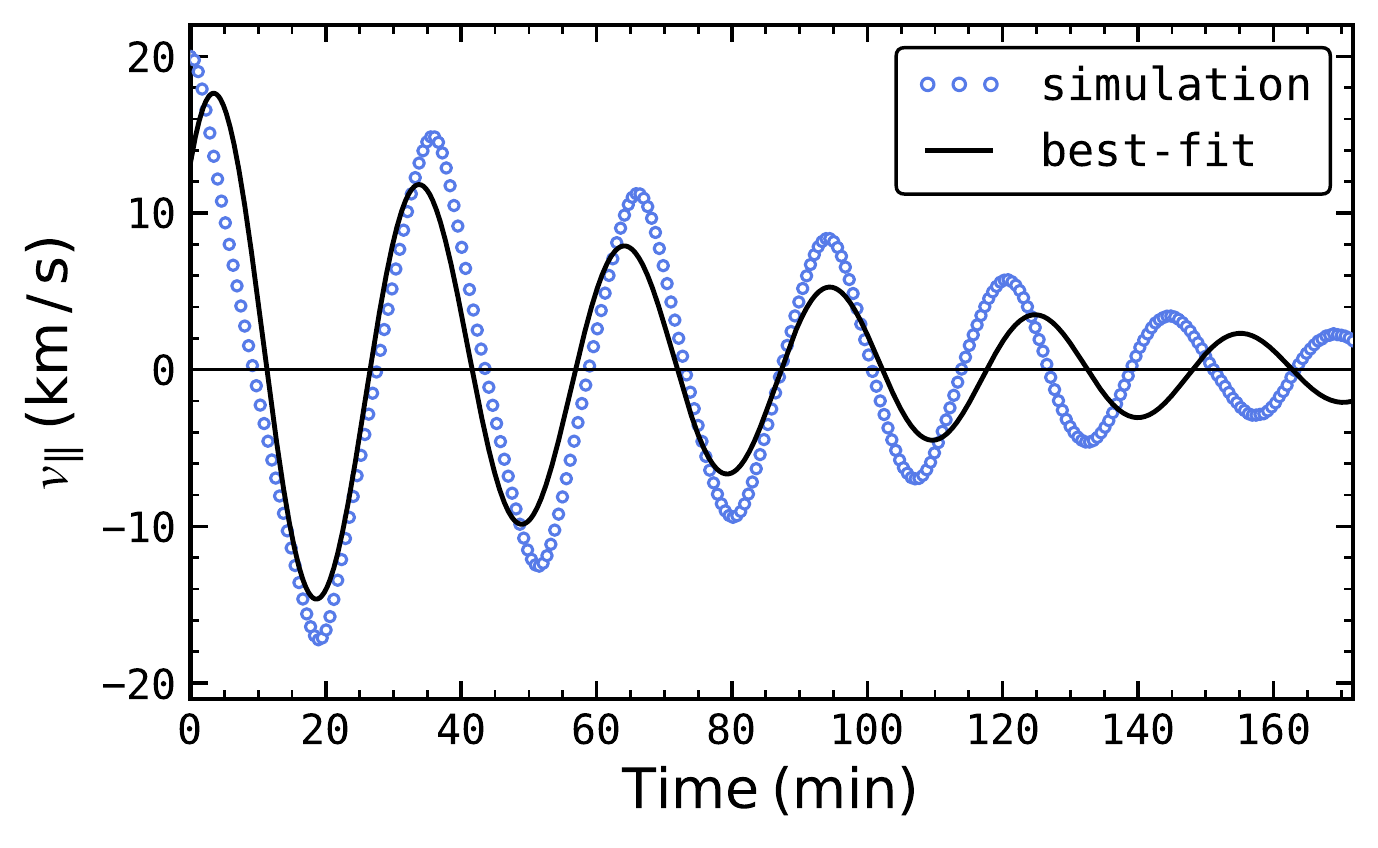}
    \caption{Left: Time-distance diagram of the temperature distribution in 
	the 1D non-adiabatic case. Right: Evolution of the velocity of the 
	filament centroid in the 1D non-adiabatic case.}
    \label{fig:fit0-1d}    %%%Fig 5
\end{figure}
%-----------------------------------------------------------------------
After the same velocity perturbation is imposed on the filament, the filament 
thread starts to oscillate. The time-distance diagram of the temperature 
distribution is displayed in the left panel of Figure \ref{fig:fit0-1d}. The 
evolution of the velocity at the filament center is shown in the right panel 
as the the blue circles. Similar to the 2D case, the velocity evolution is 
also fitted with a damped sine function the same as Equation (\ref{eq:fit0}) 
with the least-square method. It is revealed that the oscillation period is 30 
minutes and the decay time is 76 minutes. In this case, the variation of the 
period is more obvious since the deviation from the fitted curve with a fixed 
period becomes larger and larger. 

\subsection{1D and 2D adiabatic cases}

For comparison, we also numerically simulated the adiabatic cases in both 1D
and 2D. Since the evolutions are similar, the details are not described here. 
However, the fitting results are presented as follows: In the 1D adiabatic 
case, the oscillation period is $P$=37 minutes, and it is almost decayless; In 
the 2D adiabatic case, the oscillation period is $P$=44 minutes, and the decay 
time is $\tau$=211 minutes.

\section{Discussions} \label{sec:dis}

Based on the analysis of 196 filament longitudinal oscillation events observed 
in 2014, \citet{luna18} found that the ratio of the damping time to the period 
$\tau/P$ can be as small as 0.6. Several factors may contribute to the 
damping. The primary one is the non-adiabatic processes including radiation and 
heat conduction, as investigated via simulations by \citet{zhan12}. However, 
their results indicate that radiation and heat conduction are not enough to 
account for the observed damping, and other factors should play a role as well. 
One possible candidate is the mass change. According to \citet{Luna2012b}, 
the mass accretion due to continual thermal condensation would speed up the 
damping. This is understandable since increasing mass leads to deceasing 
velocity in order to conserve the total momentum. Interestingly, according to 
\citet{Zhang2013}, mass drainage would also lead to stronger damping. They 
showed that when the amplitude of the filament longitudinal oscillation is too 
large, part of the filament material drains down across the apex of the 
magnetic dip. The mass drainage takes away part of the mechanical energy of 
the filament, leading to stronger damping as well. The second additional 
candidate is the thread-thread interaction. As demonstrated by 
\citet{Zhou2018}, when there are two dips (hence two filament threads) along 
one magnetic field line, the two threads exchange kinetic energy, leading to 
weaker decay for one thread and stronger decay for the other. The third 
additional candidate is the deformation of 
the magnetic field lines, which would generate kink waves propagating outward. 
In this case, the oscillation energy is taken away via wave leakage. 
The significant feature of the filament longitudinal oscillations in our 2D 
non-adiabatic case is the rapid decay, where the decay time $\tau$ is only 
0.7 times the oscillation period $P$, i.e., $\tau=0.7P$, in contrast to 
$\tau=2.5P$ in the 1D non-adiabatic case \citep{zhan12}. Since there are no 
significant mass change and thread-thread interaction here, we discuss how the 
non-adiabatic processes and wave leakage affect the damping.

\subsection{Understanding the damping due to non-adiabatic processes}

%-----------------------------------------------------------------------
\begin{figure*}[ht!]
    \centering
    \gridline{
        \fig{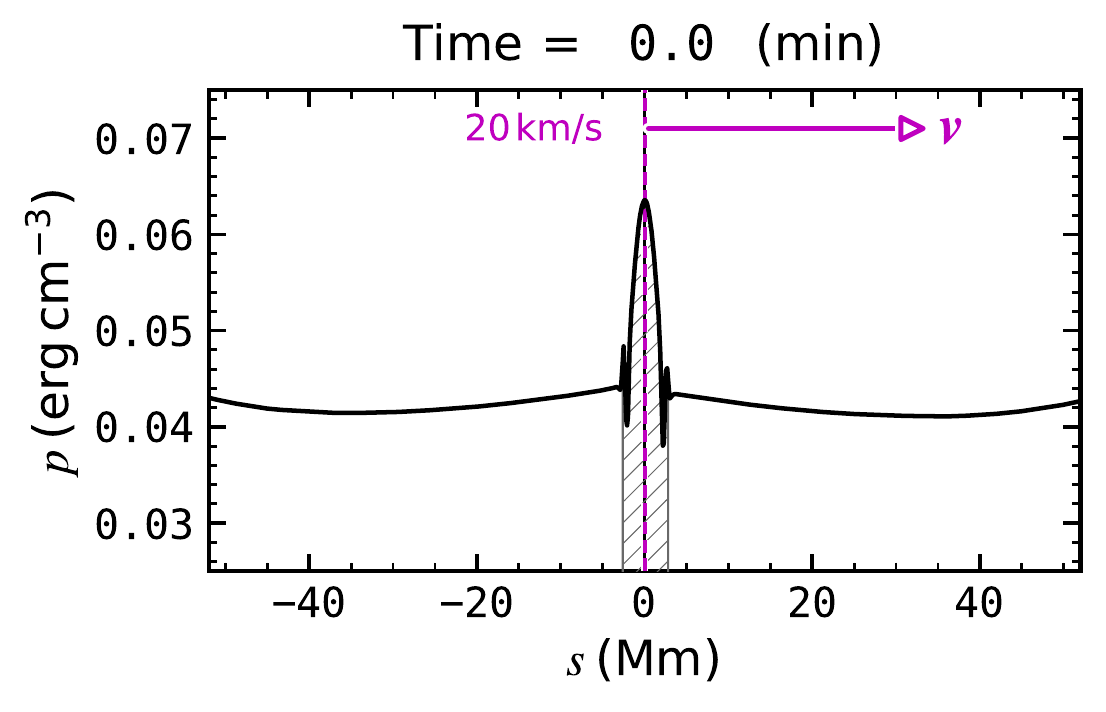}{0.45\textwidth}{}
        \fig{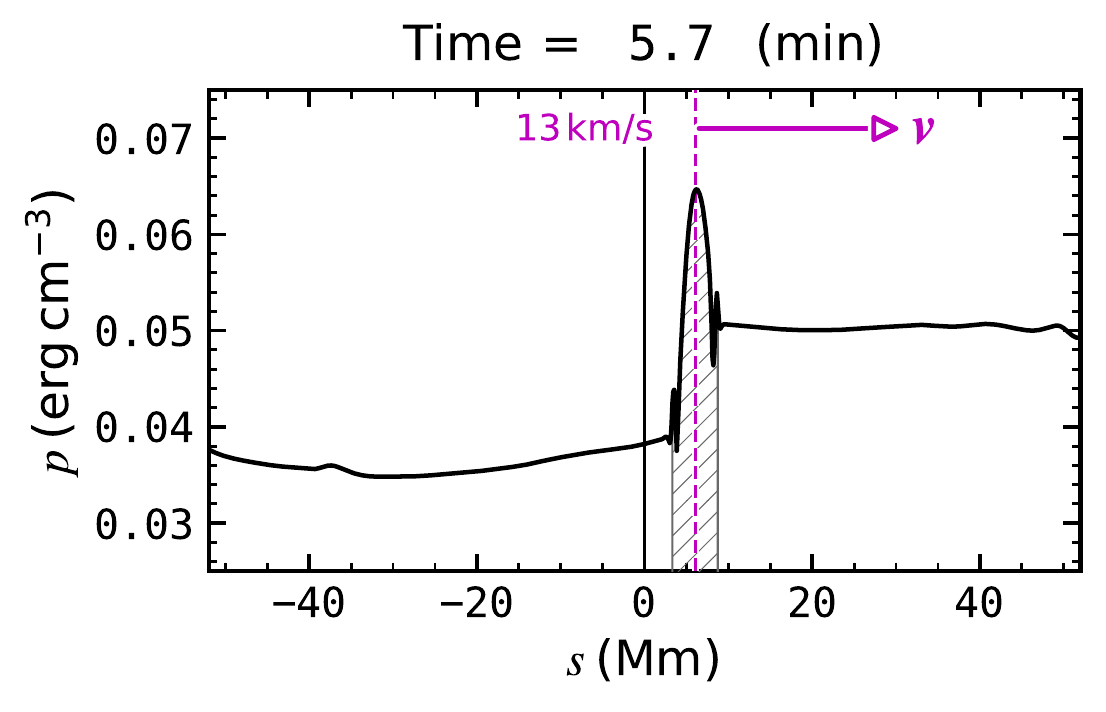}{0.45\textwidth}{}
        }
    \gridline{
        \fig{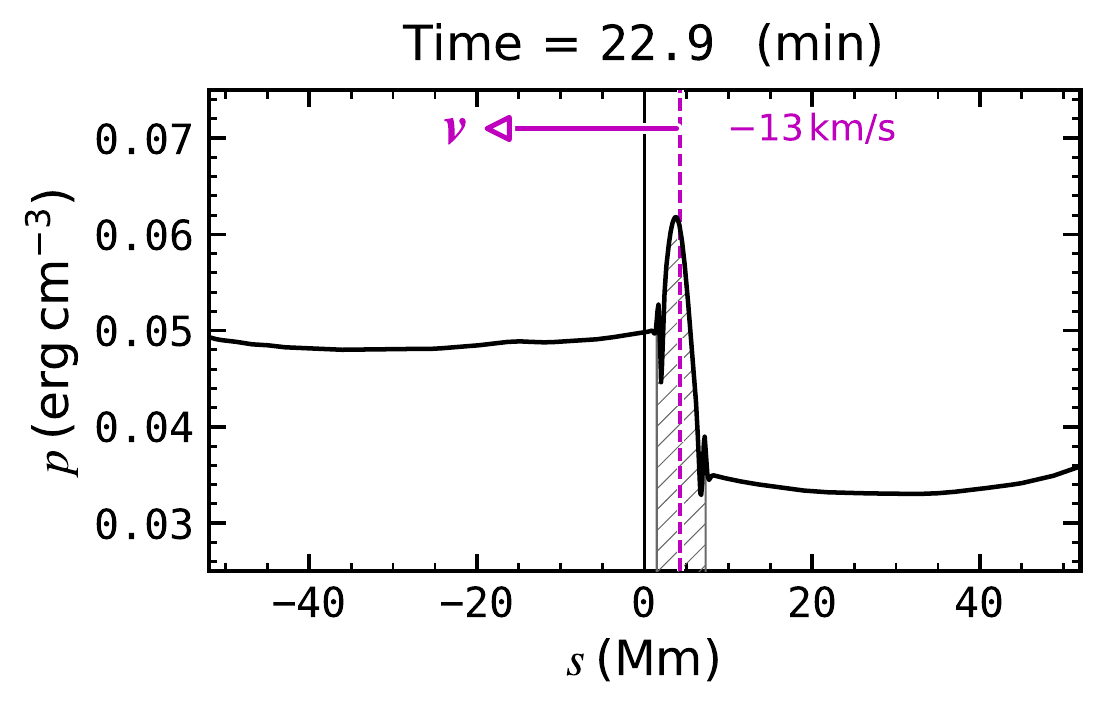}{0.45\textwidth}{}
        \fig{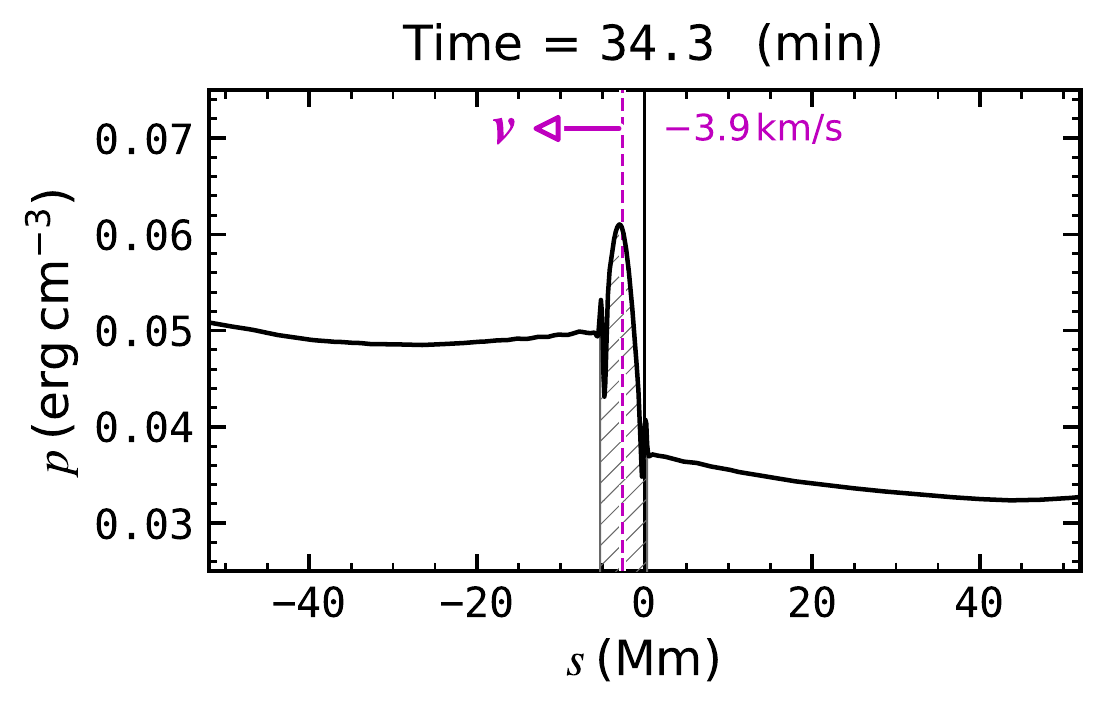}{0.45\textwidth}{}
        }
    \caption{Snapshots of the gas pressure distribution along the magnetic 
	field line crossing the filament centroid, where the thatched areas 
	corresponds to the filament, and the magenta arrows indicate the 
	velocity of the filament.}
    \label{fig:p}     %% Fig 6
\end{figure*}

From the energy point of view, it is straightforward to understand how the 
filament longitudinal oscillations decay faster because of radiation and heat 
conduction \citep[see][for details]{Zhang2013}. It is simply because radiation 
and heat conduction take away the energy of the oscillating filament. In this 
subsection, we try to understand the damping process from the force point of 
view.

\begin{figure}[h!]
    \centering
    \includegraphics[width=.9\textwidth]{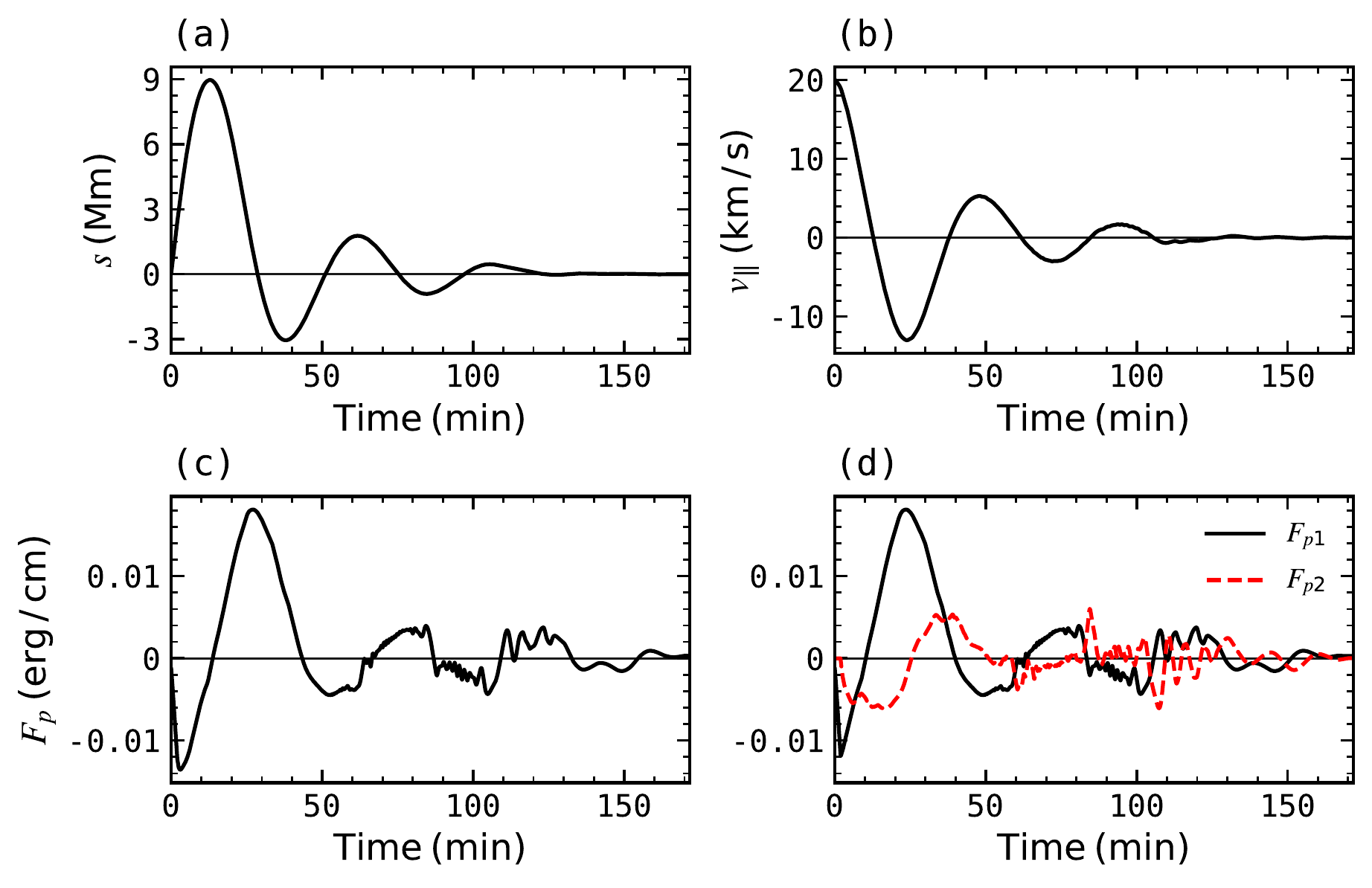}
	\caption{Evolution of several quantities in the 2D non-adiabatic case. 
	Panel (a) is for the filament displacement along the magnetic field 
	line, panel (b) is for the filament velocity along the magnetic field 
	line, panel (c) is for $F_p$, the pressure gradient force across the 
	left and right boundaries of the filament per square centimeter, and 
	panel (d) is for $F_{p1}$ and $F_{p2}$, the two decomposed parts of 
	$F_p$.}
    \label{fig:fv}    %% Fig 7
\end{figure}

For longitudinal oscillations, Lorentz force does not play a role directly, 
and gravity 
is always a restoring force. Therefore, from the dynamics point of view, the 
only possible force for the decay is the gas pressure difference between the 
left and right boundaries of the filament. Figure \ref{fig:p} displays four 
snapshots of the gas pressure distribution along the blue line marked in Figure 
\ref{fig:init}. It is seen that, different from the field-aligned gravity, the 
pressure gradient force is mostly opposite to the filament velocity, rather 
than to the filament displacement. To see this more clearly, Figure 
\ref{fig:fv} displays the evolution of the filament displacement in panel (a), 
the evolution of the filament velocity in panel (b), and the evolution of 
the pressure gradient force per square centimeter ($F_p$) in panel (c). It is 
revealed that the pressure gradient force is almost antiphase with the filament 
velocity, with a small phase difference. Considering the slight phase shift, we 
decompose the pressure gradient force $F_p$ into two parts. The first part, 
$F_{p1}$, is obtained by shifting the $F_p$ profile so that the new profile is 
exactly antiphase with $v_\parallel$. This is done by taking the maximum 
running correlation coefficient between $F_{p1}$ and $v_\parallel$. Therefore, 
$F_{p1}$ can be considered as the viscous force. $F_{p2}$ is simply the 
residual. Panel (d) of Figure \ref{fig:fv} displays the evolutions of $F_{p1}$ 
(black solid line) and $F_{p2}$ (red dashed line). It is seen that $F_{p1}$ is 
dominant, i.e., for the non-adiabatic case, the pressure gradient force acts 
mainly as a damping force, and only a minor part of it, $F_{p2}$, contributes 
slightly to the restoring force since it is antiphase with the filament 
displacement ($s$).

%-----------------------------------------------------------------------

%-----------------------------------------------------------------------
\begin{figure}[h!]
    \centering
    \includegraphics[width=.9\textwidth]{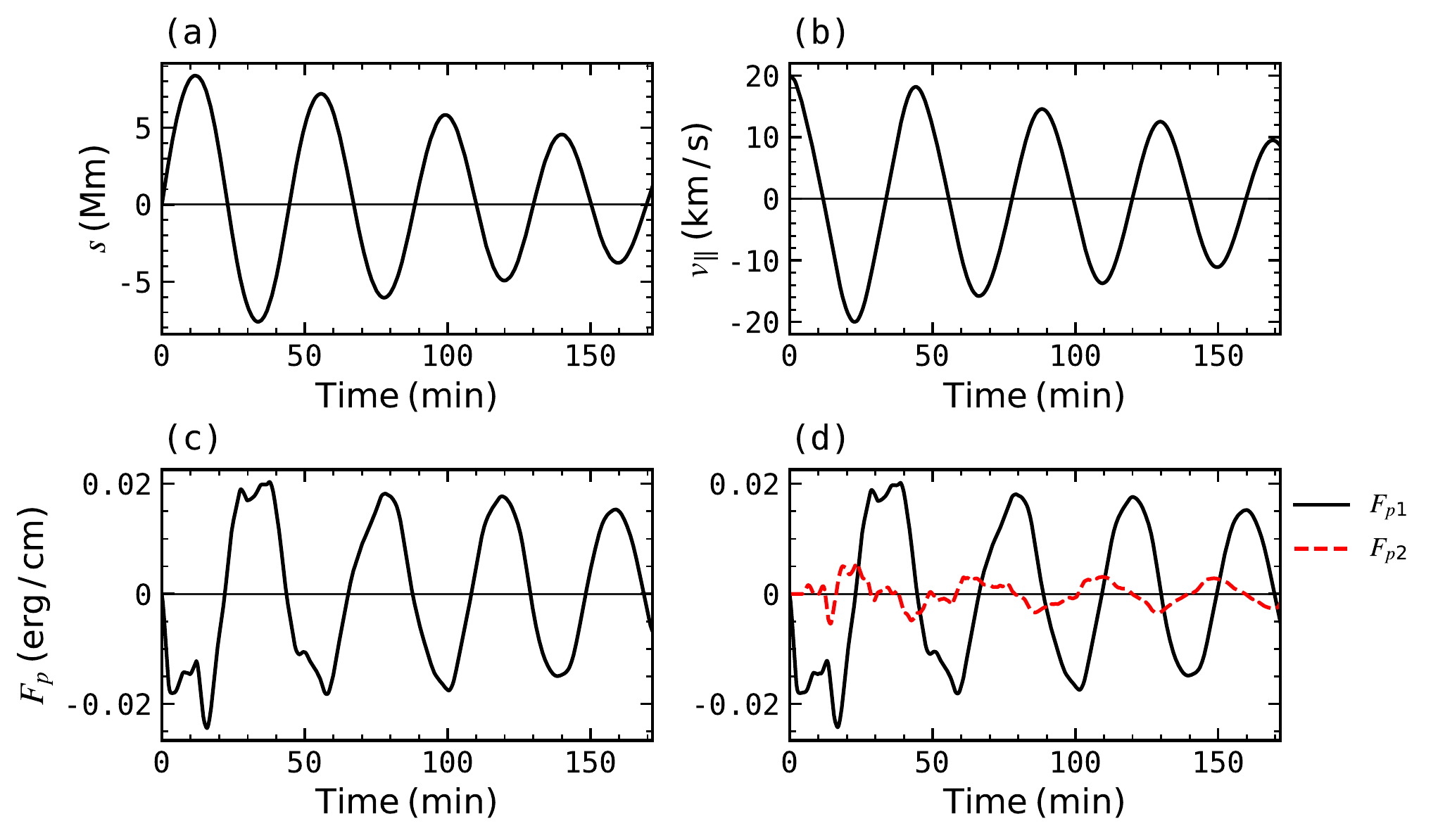}
	\caption{The same as Figure \ref{fig:fv}, but for the 2D adiabatic 
	case.}
    \label{fig:fv-adiab}	%% Fig 8
\end{figure}

For comparison, Figure \ref{fig:fv-adiab} displays the evolutions of the 
filament displacement (panel a), field-aligned velocity (panel b), gas pressure 
gradient force $F_p$ (panel c), and the decomposed two forces (panel d) in the 
2D adiabatic case. It is seen that the pressure gradient force $F_p$ is roughly 
proportional to the filament displacement $s$ with a negative coefficient, 
rather than proportional to the filament velocity as in the non-adiabatic case. 
That is to say, the pressure gradient force $F_p$ acts mainly as a restoring 
force in the 2D adiabatic case, though it is weaker than the gravity. Similar 
to that in the previous paragraph, we decompose $F_p$ into two components, 
$F_{p1}$ (black solid line) and $F_{p2}$ (red dashed line), where $F_{p1}$ is 
obtained by shifting $T_p$ to be exactly antiphase with $s$, and $F_{p2}$ is 
the residual, i.e., $F_{p2}=F_p-F_{p1}$. It is seen that in the non-adiabatic 
case, the dominant component of the pressure gradient, $F_{p1}$, contributes to 
the restoring force \citep[see][as well]{Luna2016}, and only the minor 
component, $F_{p2}$, contributes to the damping. This explains why the damping 
time is much longer in the 2D adiabatic case.

From the above analysis, it is revealed that in the adiabatic case, the gas 
pressure gradient force mainly acts as a restoring force, i.e., when the 
filament is to the right away from the equilibrium position, the coronal plasma 
on the right part is compressed with a higher gas pressure. However, when 
radiation and heat conduction are included, it is always the upstream side of 
the filament that has higher gas pressure, making the pressure gradient force 
almost a viscous force. In order to understand the reason for the 
difference, we examine the evolution of the thermal parameters in the two 
coronal segments on the two sides of the filament along the magnetic flux tube. 
It is found that, as the filament moves to the right, the right-hand coronal 
segment is compressed (and the left-hand coronal segment is rarefied), leading 
to higher gas pressure and density on the right (and lower pressure and density 
on the left). As a result of the higher density, the radiative cooling is 
enhanced on the right side (and weakened on the left). As time goes on, the 
temperature decreases on the right side (and increases on the left side), 
leading to lower gas pressure on the right side soon after the filament reaches 
its rightmost position (and higher gas pressure on the left side). 
Consequently, the pressure gradient force mainly acts as a viscous force. Note 
that radiation and heat conduction are important since the 
timescales of radiation and heat conduction in the corona, tens of minutes, 
are comparable to the period of the filament longitudinal oscillations. It is 
expected that they are much less important for filament transverse oscillations 
since the period of the latter is several times shorter.

\subsection{Damping due to wave leakage}

Comparing the 2D non-adiabatic and 1D non-adiabatic cases, it is revealed that 
the decay time in the 2D case is only half that in the 1D case. It is seen 
that the 2D filament longitudinal oscillations decay faster than the 1D 
oscillations. Such a difference is also seen between the 2D adiabatic and 1D 
adiabatic cases, where the latter is decayless and the former has a decay 
time of 76 minutes. We conjecture that the additional energy loss mechanism of 
the 2D effect is wave leakage. The reason is that the plasma $\delta$ in our 
paper, i.e., the ratio of gravity to Lorentz force, is close to unity, so the 
longitudinal oscillations of the filament would deform the magnetic field, and 
the deforming magnetic field would generate transverse waves, which then 
propagate away from the filament. 

%-----------------------------------------------------------------------
\begin{figure}[ht!]
    \includegraphics[width=0.8\textwidth]{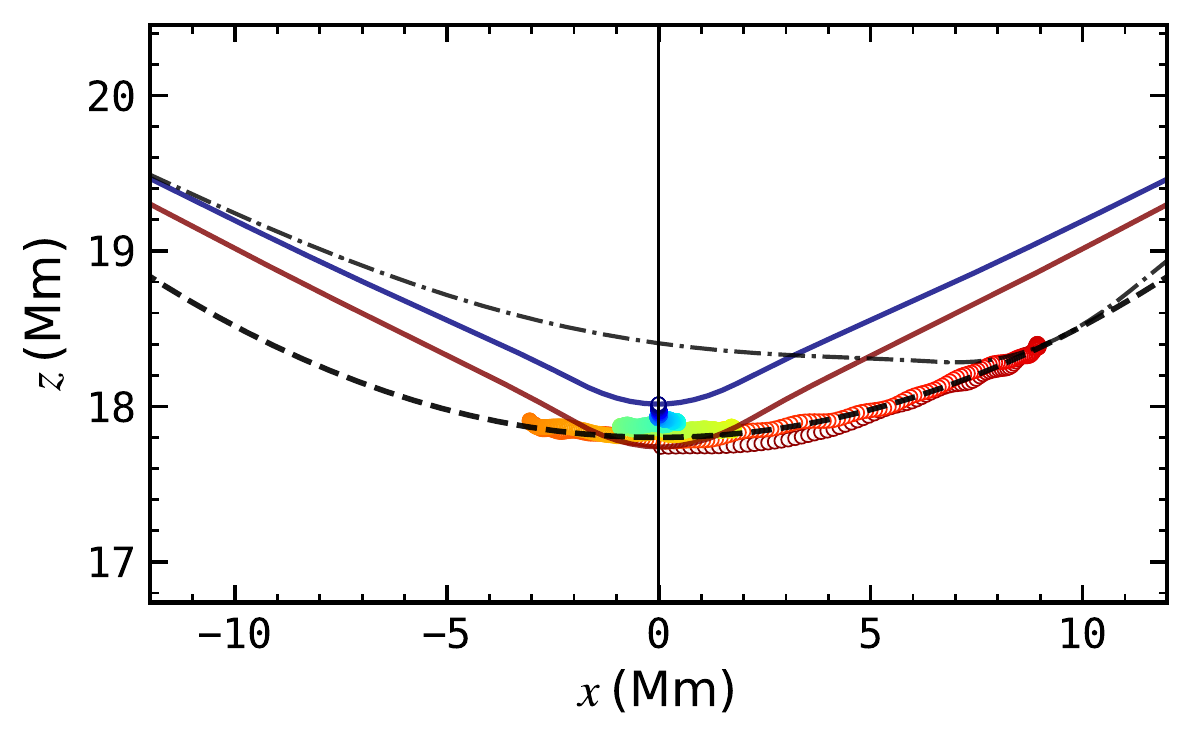}
	\caption{The trajectory of the filament centroid ({\it colored 
	circles}) and the magnetic field lines across the filament centroid, 
	where the red line corresponds to the initial state, the 
	dash-dotted line corresponds to the moment when the filament reaches 
	its rightmost position, and the blue line 
	corresponds to $t=171.75$ minutes. The dashed line is a fitted 
	circular arc with a radius of 70 Mm. Note that the horizontal and 
	vertical axes are not to scale.}
    \label{fig:traj}  	% Fig 9
\end{figure}
%-----------------------------------------------------------------------
In order to confirm it, we plot the evolution of the magnetic field line 
threading the filament centroid at the initial ({\it red line}) and the final 
({\it blue line}) times in Figure \ref{fig:traj}. In this figure, the 
trajectory of the filament centroid is overplotted as small colored circles, 
where the color coding from red to blue indicates the time elapse. We can see 
that, as the filament moves, the magnetic field line is deformed by at least 
$\sim$0.5 Mm.

%-----------------------------------------------------------------------
\begin{figure}[ht!]
    \centering
    \includegraphics[width=.8\textwidth]{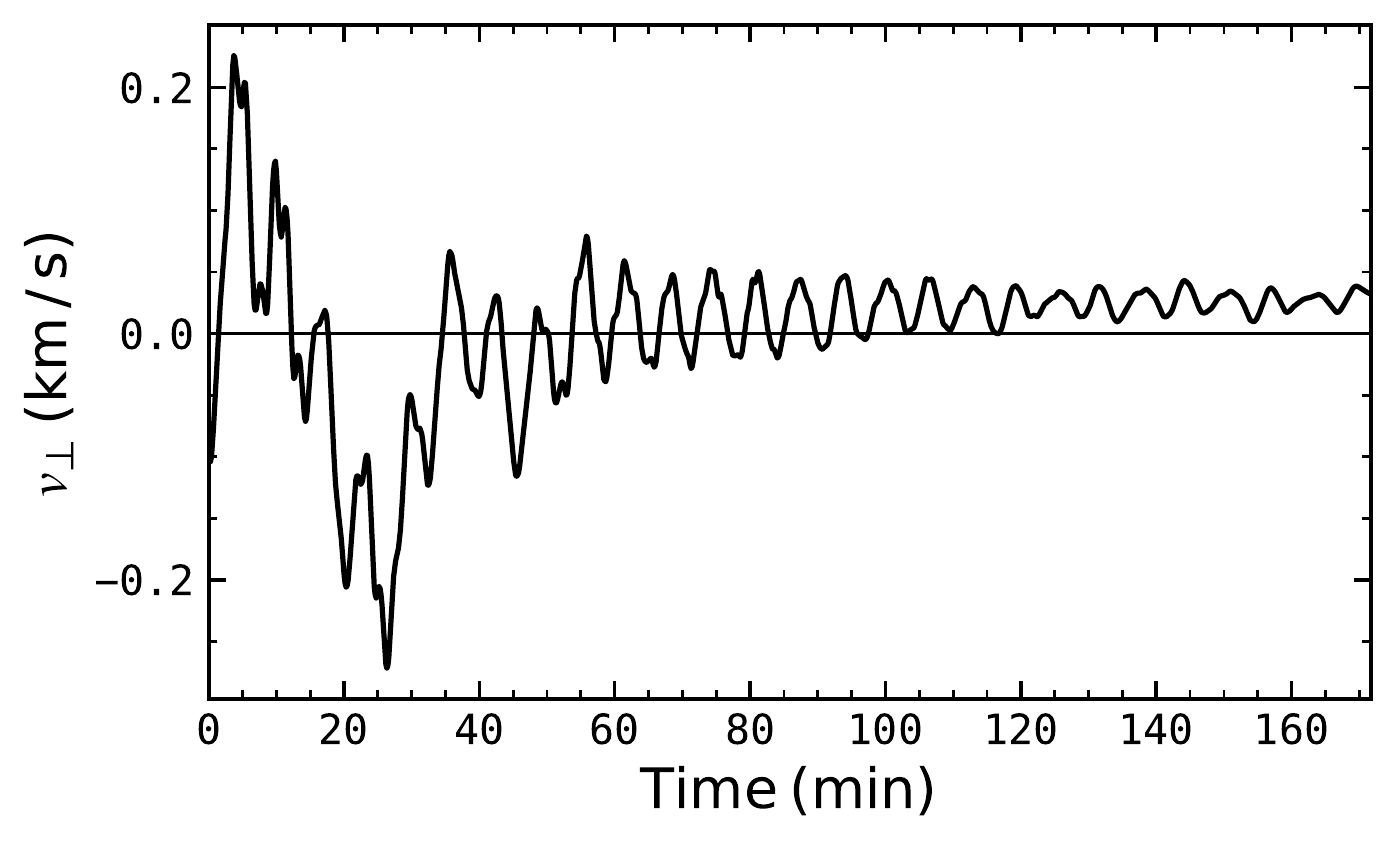}
    \caption{Evolution of the vertical velocity of the filament, showing a decayed 
	transverse oscillation. }
    \label{fig:vz} 
\end{figure}
%-----------------------------------------------------------------------
As expected, such deformation would excite filament transverse oscillations.
Figure \ref{fig:vz} displays the temporal evolution of the vertical velocity 
of the filament centroid in the 2D non-adiabatic case. It is seen that 
short-period oscillations are superposed on a quickly damped oscillations with 
a larger period. The short-period oscillation has a period of 6.38 minutes, 
which is in the typical range of filament transverse oscillations  
\citep{trip09,hill13}, whereas the quickly-damped oscillation has a period of 
25 minutes, which is about half of the filament longitudinal oscillation. The 
half period is expected when a filament is oscillating along a dipped magnetic 
field. Note that the amplitude of transverse oscillation induced by the 
longitudinal oscillation is only 0.2 \kms, which is in the lower range detected 
by the {\it Hinode} satellite \citep{hill13}, and much smaller than the 
velocity amplitude of filament transverse oscillations indcued by coronal 
waves, e.g., 8.8 \kms\ in \citet{liu12} and 65--89 \kms\ in \citet{shen14}.

%-----------------------------------------------------------------------
\begin{figure}[ht!]
    \centering
    \includegraphics[width=.9\textwidth]{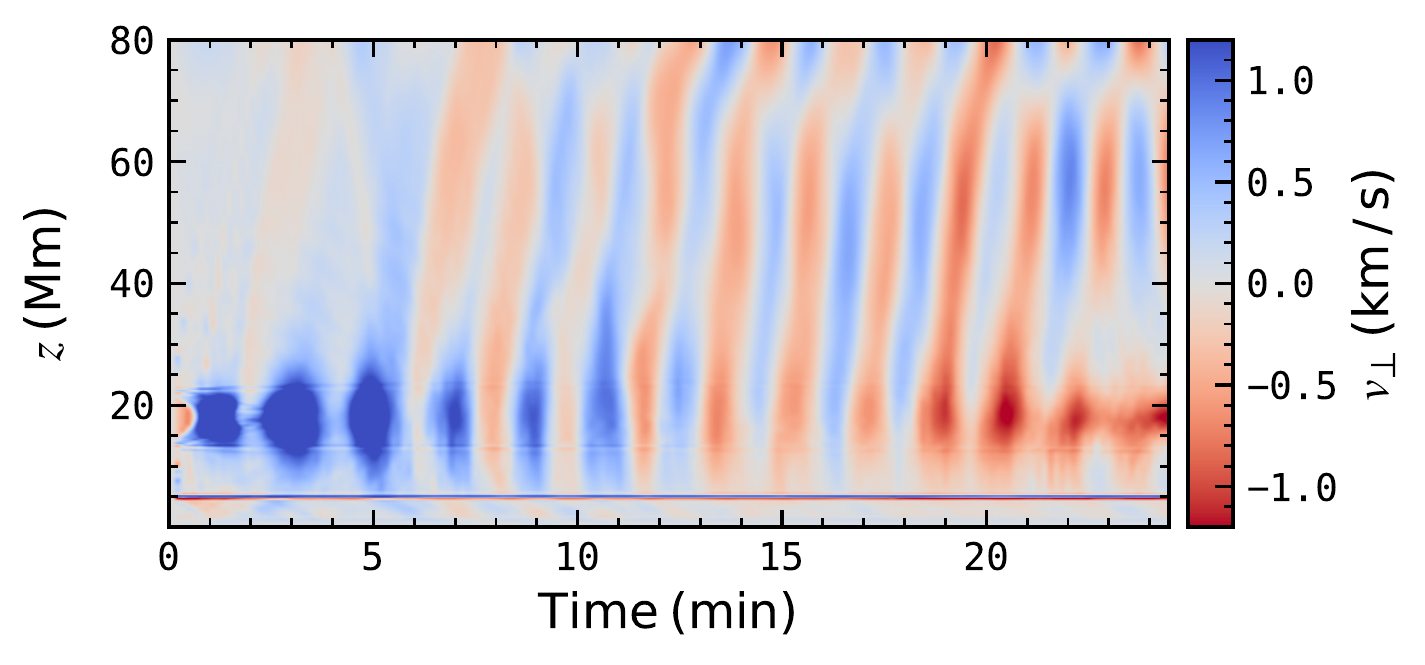}
    \caption{Time-distance diagram of the vertical velocity ($v_\perp$) along 
	the $z$-axis at $x=0$.}
    \label{fig:v-z} 	%% Fig 11
\end{figure}
%-----------------------------------------------------------------------
As the magnetic field line across the filament oscillates vertically, it is 
anticipated to see fast-mode waves being excited. To demonstrate the fast-mode 
wave generation, we plot the time evolution of the $v_z$ distribution along 
the $z$-axis in Figure \ref{fig:v-z}, where blue color means upward velocity 
and red color means downward velocity. It is seen that quasi-periodic waves 
propagate both in the upward and the downward directions. The period of the 
wave train is calculated to be $\sim$6.38 minutes based on wavelet analysis, 
which is almost identical to that of the filament transverse oscillation. The 
propagation speed of the fast-mode waves as indicated by the slope of the 
oblique ridges is measured to be $810\pm 60$ km s$^{-1}$, the typical fast-mode 
wave speed in the background corona.

\begin{figure}[ht!]
    \centering
    \includegraphics[width=.8\textwidth]{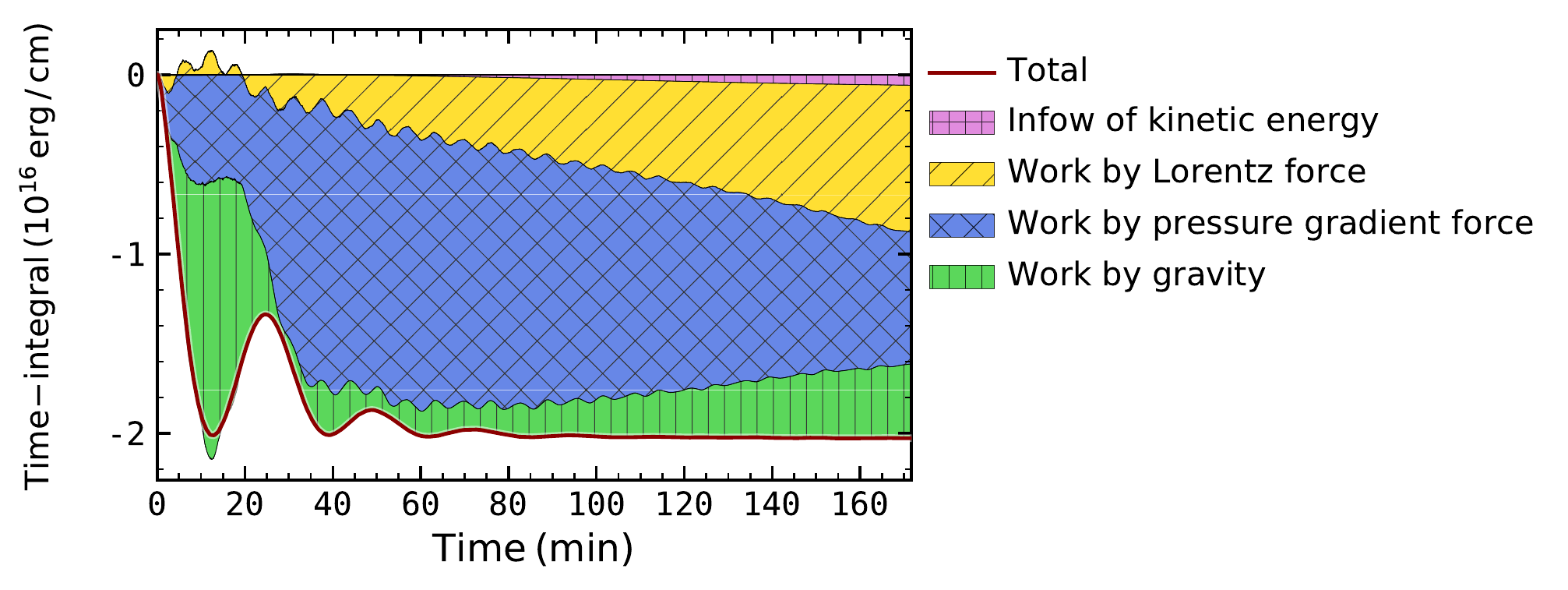}
    \caption{Time-integral of the energy loss from the fixed region indicated 
	by the pink box in Figure \ref{fig:evo}.}
    \label{fig:Uk}	%% Fig 12
\end{figure}

In order to quantify how much energy is taken away by the fast-mode MHD waves, 
we select a fixed rectangular region around the oscillating filament as 
indicated by the pink box in Figure \ref{fig:evo}. Note that the box is large 
enough so that the filament is always inside the box. Within the box, 
we calculate the kinetic energy convected into the box ($E_{\ek} = 
-\oint_{L} \ek \boldsymbol{v} \cdot \boldsymbol{n} d l$, where $\ek$ is the 
kinetic energy density, $\boldsymbol{n}$ is the normal vector, and $l$ is the 
length along the box boundary), the work done by Lorentz force 
($E_B=\iint_{\Sigma} \boldsymbol{v} \cdot (\boldsymbol{j}\times \boldsymbol{B}) 
d \sigma$, where the integral is done over the whole area of the box), and 
the work done by gas pressure ($E_p=-\iint_{\Sigma} \boldsymbol{v} 
\cdot \nabla p d\sigma$, where the integral is done over the whole area of the 
box). We also calculate the work done by gravity inside the box 
($E_g= \iint_{\Sigma} \boldsymbol{v} \cdot \boldsymbol{F}_g d \sigma$, where 
the integral is done over the whole area of the box). The evolution of the 
four parts of the accumulating energy is displayed in Figure \ref{fig:Uk}, 
where the pink area corresponds to the influx of the kinetic energy, the 
yellow area corresponds to the work done by Lorentz force, the blue area 
represents the work done by gas pressure, the green area represents the 
work done by gravity, and the red line corresponds to the sum of all the 
terms. It is seen that near the end of the simulation when the oscillation is 
almost completely attenuated, the work done by Lorentz force and gas pressure 
is dominantly negative, meaning that most of the initial kinetic energy of the 
filament longitudinal oscillation is lost by the work done by Lorentz force 
and gas pressure. Since fast-mode magnetoacoustic waves are due to the 
synchronous action of the Lorentz force and the gas pressure, it further 
implies that the major part of the initial kinetic energy of the filament is 
taken away by wave leakage.

%-----------------------------------------------------------------------
%-----------------------------------------------------------------------

Wave leakage was proposed to be able to lose the energy of an oscillating 
coronal loop to the surroundings 
\citep{call86}. Further simulations \citep{brad05, selw06} and mode analysis 
\citep{diaz06, verw06} indicated that in a 2D slab model, wave energy cannot 
be trapped, and wave leakage can efficiently attenuate the coronal loop 
oscillations. \citet{terr06}, however, pointed out that with a flux tube in a 
3D magnetic configuration, wave leakage is not effective anymore, and 
resonant absorption becomes the dominant damping mechanism. The difference 
between the 2D slab model and the 3D flux tube model can be understood in an 
intuitional way: In the 2D slab model, whenever the coronal loop oscillates 
vertically, all the background magnetic field lines are pushed or pulled to 
oscillate. Hence, it is nor surprising that wave leakage is an efficient 
mechanism for the energy loss. On the other hand, in the 3D model where the 
flux tube is embedded in a magnetic field roughly aligned with the flux 
tube, when the flux tube oscillates, the ambient field lines may be pushed 
aside because of the third degree of freedom, and are not necessary to 
oscillate following the oscillating flux tube. That is to say, the oscillating 
slab in the 2D model is similar to a piston, which is not the case in 3D if 
the background magnetic field is roughly parallel to the flux tube. It is 
pointed out here that, however, if the background magnetic field is 
quasi-perpendicular to the flux tube, the oscillating flux tube would serve as 
a piston even in the 3D case. In this situation, all the ambient magnetic 
field lines would be pushed or pulled to oscillate, leading to significant 
wave leakage. The magnetic configuration of solar filaments is exactly like 
this, i.e., a sheared (and maybe twisted) core field overlain by unsheared 
envelope magnetic field \citep{chen11, pare14}. As a result, when the 
filament, which is held by the core field, oscillates vertically, it would 
generate kink motions of the envelope field, leading to wave leakage.

\subsection{Validity of the pendulum model}

For filament longitudinal oscillations, it is often believed that the 
restoring force is the field-aligned component of gravity. Therefore, pendulum 
model was used to relate the oscillation period to the curvature radius of 
the dipped magnetic field, as confirmed by the 1D hydrodynamic simulations 
\citep{Luna2012b, zhan12}. In order to verify the pendulum model in 2D where 
magnetic field may react to the field-aligned motion of the filament, 
\citet{Luna2016} performed 2D MHD simulations, where the magnetic field around 
the filament is as weak as 10 G. It was found that the pendulum model still 
holds and there is no strong coupling between longitudinal and transverse 
oscillations of the filament. It is noted that, according to \citet{Zhou2018}, 
whether the magnetic field may significantly react or not depends on the 
dimensionless parameter $\delta$, i.e., the gravity to Lorentz force ratio, 
rather than the plasma $\beta$, i.e., the gas to magnetic pressure ratio. We 
checked the plasma $\delta$ in \citet{Luna2016}, and found that it is about 
0.2, i.e., its gravity is several times smaller than the Lorentz force. 
Therefore, it is natural to see that the magnetic field deformation was small 
and the filament longitudinal oscillation did not excite significant transverse 
oscillation in their simulations.

In this paper, we extended the plasma $\delta$ regime to near unity. We found 
that fast-mode wave trains are excited and propagate upward and downward away 
from the oscillating filament, taking away a significant portion of the 
oscillation energy. With radiation and heat conduction considered, the decay 
time of the longitudinal oscillation is 76 minutes in the 1D simulation, and 
becomes 38 minutes in the 2D simulations. The deformation of the magnetic 
field is not trivial, which is up to 0.5 Mm as evidenced in Figure 
\ref{fig:traj}. Because of the deformation of the magnetic field, even the 
oscillation period is different between the 2D and 1D cases. As seen from 
Figure \ref{fig:traj}, as the filament moves to the right, the corresponding 
portion of the magnetic field is pressed to be flatter as indicated by 
the dash-dotted line in Figure \ref{fig:traj}, which results in a 
smaller component of the gravity along the field line. As a result, the 
corresponding period in the 2D case becomes longer. This explains why the 
oscillation period in the 1D non-adiabatic case is 30 minutes, whereas the 
period becomes 49 minutes in the 2D non-adiabatic case. The relative error is 
more than 50\%. Since the curvature radius of the magnetic dip is proportional 
to the square of the oscillation period according to the pendulum model, the 
relative error of the curvature radius would be 100\%. Because the gravity is 
comparable to the Lorentz force, the field line is strongly curved near the 
filament and becomes almost straight further away as evidenced by the red 
line in Figure \ref{fig:traj}. Such an extremely non-uniform curvature 
distribution accounts for the decreasing period as the oscillation attenuates 
seen in the right panels of Figures \ref{fig:Te-s} and \ref{fig:fit0-1d}.

It is interesting to notice that, although the magnetic field line across the 
filament centroid is not circular, the actual trajectory of the filament is
almost circular. As revealed by the colored circles in Figure \ref{fig:traj}, 
the trajectory fits into a circular arc ({\it dashed line}) very well.

%%%%%%%%%%%%%%%%%%%%%%%%%%%%%%%%%%%%%%%%%%%%%%%%
\section{Summary}\label{sec:sum}

In this paper, we investigated the filament longitudinal oscillations in
the weak magnetic field regime. The main results are summarized as follows:

(1) Our simulations verified the suggestion proposed by \citet{Zhou2018},
i.e., whether the dense plasma of a filament may modify the magnetic
field is not determined by the plasma $\beta$, it is determined by the
gravity to Lorentz force ratio $\delta={{\rho gL}\over {B^2/2\mu_0}}$.
In our case where $\beta$ is small but $\delta$ is close to unity, the
magnetic field is substantially changed by the oscillating filament.
That is, low plasma $\beta$ does not guarantee that the magnetic field
lines are not changed by the filament gravity. In the high $\delta$
case as in this paper, tha application of the pendulum model would lead to an 
error of $\sim$100\% in estimating the curvature radius of the dipped magnetic 
field.

(2) In the framework of 2D simulations, the inclusion of heat conduction
and radiation significantly reduce the decay time from 211 minutes in
the adiabatic case to 34 minutes in the non-adiabatic case, implying
that the non-adiabatic processes are the primary agent that dissipates
the kinetic energy of the filament.

(3) With heat conduction and radiation being included, the decay time is
remarkably reduced from 113 minutes in the 1D case to 34 minutes in the
2D case. It is found that the filament longitudinal oscillations excite
transverse fast-mode waves, and the wave leakage is one important agent
that can lose the kinetic energy of the filament.

%%%%%%%%%%%%%%%%%%%%%%%%%%%%%%%%%%%%%%%%%%%%%%%%%%%%%%%%%%%%%%%%%%%%%%%%%%%%%%
\acknowledgments

This research was supported by the Chinese foundations (NSFC 11533005,
11961131002, and U1731241) and Jiangsu 333 Project (No. BRA2017359). The 
numerical calculations in this paper were performed on the cluster system in 
the High Performance Computing Center (HPCC) of Nanjing University. PFC
thanks ISSI and ISSI-Beijing for supporting team meetings on solar filaments.

%%%%%%%%%%%%%%%%%%%%%%%%%%%%%%%%%%%%%%%%%%%%%%%%%%%%%%%%%%%%%%%%%%%%%%%%%%%%%%
\bibliographystyle{aasjournal}

\bibliography{ms}
% \nocite{*}

\end{document}